\documentclass[acmlarge]{acmart}
\AtBeginDocument{%
  }

\acmJournal{TOMM}
\acmVolume{xx}
\acmNumber{xx}
\acmArticle{Under Review}
\usepackage{algorithmic}
\usepackage{graphicx}
\usepackage{textcomp}
\usepackage{booktabs}
\usepackage{subfigure}

\def\eg{\textit{e.g.}}
\def\ie{\textit{i.e.}}

\def\etc{\textit{etc}}

\begin{document}

\title{EEG-based Cognitive Load Estimation of Acoustic Parameters for Data Sonification}

\author{Gulshan Sharma}
\orcid{0000-0002-5332-7256}
\affiliation{%
 \institution{Indian Institute of Technology Ropar}
  \country{India}}

\author{Surbhi Madan}
\orcid{0009-0000-3774-8117}
\affiliation{%
 \institution{Indian Institute of Technology Ropar}
  \country{India}}

\author{Maneesh Bilalpur}
\orcid{0009-0005-6454-6349}
\affiliation{%
 \institution{University of Pittsburgh}
  \country{USA}}

\author{Abhinav Dhall}
\orcid{0000-0002-2230-1440}
\affiliation{%
 \institution{Flinders University}
 \country{Australia, and}
  \institution{Indian Institute of Technology Ropar}
 \country{India}}

\author{Ramanathan Subramanian}
\orcid{0000-0001-9441-7074}
\affiliation{%
 \institution{University of Canberra}
 \country{Australia}}

\renewcommand{\shortauthors}{Sharma et al.}

\begin{abstract}
Sonification is a data visualization technique which expresses data attributes via \textit{\textbf{psychoacoustic parameters}}, which are non-speech audio signals used to convey information. This paper investigates the binary estimation of cognitive load induced by psychoacoustic parameters conveying the focus level of an astronomical image via Electroencephalogram (EEG) embeddings. Employing machine learning and deep learning methodologies, we demonstrate that EEG signals are reliable for (a) binary estimation of cognitive load, (b) isolating easy vs difficult visual-to-auditory perceptual mappings, and (c) capturing perceptual similarities among psychoacoustic parameters. Our key findings reveal that (1) EEG embeddings can reliably measure cognitive load, achieving a peak F1-score of 0.98; (2) Extreme focus levels are easier to detect via auditory mappings than intermediate ones, and (3) psychoacoustic parameters inducing comparable cognitive load levels tend to generate similar EEG encodings.
\end{abstract}




\maketitle
\section{Introduction}\label{sec:introduction}
Data visualization is a powerful tool for communicating information in an understandable and interpretable manner. By encoding data via visual attributes such as color, shape, and size, one can easily perceive patterns and relationships within the data. This allows for users to make informed decisions and gain instant insights. Typically, traditional approaches to data visualization are limited to visual means~\cite{Bilalpur,Ferguson18,Parekh18}. These methods apply graphical representations which enable the user to interpret the semantic structure of qualitative and quantitative data. Some of the commonly applied graphical representations that enable descriptive and inferential data analysis are charts, graphs, and histograms \cite{chart_book}. With the advent of \textit{big data}, these displays have become ubiquitous in a variety of domains. However, exclusive dependence on visual representations can make data understanding difficult or inaccessible for users with visual impairments. Thus, alternative sensory mechanisms such as auditory, tactile, and olfactory should be considered while developing effective data visualization tools~\cite{Auditory,Tactile}, which can increase accessibility and provide profound insights into complex data.

Auditory display is a term used to denote the process of communicating information to the user through sound or music~\cite{aud_dis}. The goal is to exploit the human auditory system's ability to interpret auditory parameters and process sounds in a wide range of situations. Auditory displays have become more popular recently as they are capable of representing complex temporal transitions. They are also well-suited for use in high-stress environments where timely and accurate notifications are critical. In medicine, auditory displays can alert medical professionals to critical events, such as changes in a patient's vital signs or the onset of an adverse event~\cite{aud_med}. Similarly, auditory displays in sports can provide immediate feedback on an athlete's performance, allowing for rapid improvements~\cite{aud_sport}. In addition to medical and sports, auditory displays can also be applied in a variety of other fields, such as aviation, transportation, and entertainment \cite{aud_app}. 

Data sonification is a specific type of auditory display which uses non-speech audio parameters such as pitch, tone, and rhythm to represent data~\cite{Hartmann}, to enable users to understand data patterns, trends, and relationships~\cite{Kramer}. Sonification has applications in several domains like scientific visualization, navigation, data analysis, and is especially useful for users who have difficulty working with visual data displays. For instance, it can help visually-impaired users to move around complex environments~\cite{icmi_soni_module}, enhance understanding of scientific data~\cite{soni_literacy}, and examine patterns and trends in multi-dimensional data~\cite{fish}. 

It is also important to choose an effective sonification design that takes into account the target audience and the desired outcome~\cite{soni_dj}. \textit{E.g.}, sonification designed for a medical setting may require a different approach to one designed for sports analysis. Moreover, user studies are critical for evaluating effectiveness of these designs~\cite{soni_user}. Via user studies, designers can gather feedback on the sonification design which can then be applied for design refinement. In user studies, cognitive load (CL) is a critical consideration~\cite{Bilalpur}. Under high CL, users may experience difficulty in understanding and retaining the presented information. Due to limited mental resources to store information, users can get overwhelmed easily, leading to frustration and decreased mental performance. Therefore, it is crucial to ensure that sonification design is optimized to induce minimal CL~\cite{Bilalpur,soni_cog}.

Cognitive load can be measured through subjective self-report, behavioral, and physiological correlates such as Electroencephalogram (EEG) responses. Subjective self-reports can be captured via questionnaires, while behavioral measures may be obtained by recording explicit performance changes during problem-solving tasks. Differently, EEG-based procedures have been also developed to measure CL~\cite{cogeeg1,cogeeg2,cogeeg3,edu_eeg_cl}, where brain activity changes associated with mental effort are examined. Event-related potentials (ERP), frequency analysis, source localization, and neurofeedback are some common EEG-based CL measurement methods~\cite{erp,fa,sl,eegdelta}.

This paper builds upon prior studies~\cite{9,Bilalpur} that (a) examine the use of psychoacoustic parameters to convey visual information, namely, the focus level of an astronomical image, and (b) automatically estimate CL induced by psychoacoustic parameters such as pitch, roughness, noise, plus their combination. We extend these studies to (1) efficiently estimate CL employing multiple machine plus deep learning architectures, and features; (2) distinguish between easy vs difficult auditory mappings, and (3) estimate similarities among EEG embeddings induced by the above parameters. 

Our research contributions are as summarized below:
\begin{itemize}
    \item[1.] Different from recent EEG-based CL estimation studies~\cite{Sarailoo22}, which investigate CL induced by visual designs, we estimate CL induced by psychoacoustic parameters such as noise, pitch, roughness, \etc.
    \item[2.] We employ multiple learning architectures and descriptors to estimate cognitive load. A Multi-Layer Perceptron (MLP) network combining spatio-temporal embeddings achieves a peak CL classification F1-score of 0.98, substantially outperforming spatial and temporal encodings, thereby demonstrating the value of multimodal analytics.
    \item[3.] We also show via a Siamese network that similar EEG encodings are generated by psychoacoustic parameters inducing similar CL, whereas distinctive EEG encodings are generated by parameters inducing dissimilar CL.
\end{itemize}


 The rest of the paper is structured as follows. Section~\ref{Sec:RW} reviews related work, while Sec.~\ref{Sec:ED} details the experimental design. Section~\ref{Sec:EEG_FE} presents the methods applied for EEG pre-processing and feature extraction. Sections~\ref{Sec:ERA}--~\ref{Sec:Per_Sim} describe analyses of explicit and implicit (EEG) user responses, discriminability between EEG encodings corresponding to extremum vs intermediate focus-levels, and recognizing perceptual similarities among EEG encodings induced by the psychoacoustic parameters. Section~\ref{Sec:Con} presents our Discussions and Conclusions.

\section{Related Work}~\label{Sec:RW}
This section examines literature on data sonification and cognitive load estimation to illustrate the novelty of our work. 

\subsection{Data Sonification}

Kramer \textit{et al.} \cite{Kramer} define sonification as the use of non-speech audio to convey information, and discuss design challenges in effective sonification. Hermann~\textit{et al.}~\cite{Hartmann} provide an in-depth overview of data sonification and discuss design principles, evaluation methodologies and applications. Early sonification works focus on generating simple warning sounds~\cite{early_1}. Walker~\textit{et al.}~\cite{26} investigate various mappings in a process-control task environment where pressure, temperature, and size are data parameters mapped onto audio parameters such as loudness, pitch, tempo, and onset. Their findings suggest that intuitive data-to-sound mappings alone may not induce the best user experience, demanding empirical evaluation of auditory displays.

Ferguson~\textit{et al.}~\cite{9} evaluate multiple psychoacoustic parameters for communicating an astronomical image's focus level. They choose noise, pitch, roughness, sharpness, and their combinations as psychoacoustic parameters. Their efficacy for focus-level determination is evaluated by computing data-to-sound mapping accuracies. The author conclude that these parameters can be used to communicate image focus levels. Noise and roughness are found to be better for representing blur, whereas clear-sounding parameters are better for focus representation. The authors further conclude that users reliably determine the focus-level on a coarse-grained scale of 5 rather than a fine-grained scale of 10.

Ferguson~\textit{et al.}~\cite{ten} also examine how well pitch, noise, roughness, and a combination of noise and roughness can express negative attributes. The change in the perceived magnitude of the data variable conveyed via the acoustic attribute is mapped using magnitude estimation. The authors conclude that the acoustic features significantly influence auditory perception, and that an efficient sonification mapping will facilitate user perception. Bilalpur~\textit{et al.}~\cite{Bilalpur} collect explicit and implicit user responses to examine auditory perception, and examine EEG-based Event Related Potentials (ERPs) and spectral features to this end. The authors also perform CL estimation employing machine and deep learning models to achieve a peak F1-score of 0.64. 

\subsection{Cognitive Workload Estimation}
A substantial body of literature has demonstrated cognitive workload as a valuable measure of task difficulty. Previous work on CL evaluation employs indirect measures such as the frequency of learner errors, the time required to solve problems, and dual task computational models~\cite{cog_review}. However, these methods have since been replaced by self-ratings, in which users report the mental effort they expend for task completion. As per the notion of CL, learning and memory are affected by three different forms of cognitive load-- \textit{intrinsic}, \textit{extraneous} and \textit{germane}~\cite{book_cogload, cog_review}. Intrinsic CL refers to the implicit hardship involved in a task, while extraneous cognitive load refers to the factors that amplify the mental demands for processing and retaining information. Germane cognitive load refers to the mental effort needed to establish connections among pieces of information.

Physiological measurements can serve as objective markers for estimating cognitive workload, with EEG being a widely used modality in this regard~\cite{eeg_cog}. 
Zarjam \textit{et al.}~\cite{spec_eeg_CL} investigate EEG spectral properties acquired from EEG signals for CL estimation during a reading task. The authors conclude that spectral measures such as weighted mean frequency, spectral entropy, spectral edge frequency, and bandwidth can successfully differentiate load levels. Kumar~\textit{et al.}~\cite{band_eeg_CL} examine EEG power bands that best correlate with CL for arithmetic tasks involving auditory instructions. Mills \textit{et al.}~\cite{edu_eeg_cl} examine CL via EEG while learning with educational technology, to show the feasibility of collecting EEG data in unconstrained environments. Feature selection on power spectral density (PSD) EEG features is employed to evaluate CL imposed by educational multimedia to achieve 84\% accuracy. The generalizability of EEG-based CL estimation across different graphic visualization types is examined in~\cite{Parekh18}. A peak F1-score of 0.64 is achieved with a 3-layer CNN for binary (high vs low) classification of CL induced by psychoacoustic parameters in~\cite{Bilalpur}.

\subsection{Analysis of Related Work}
The literature reveals that (a) only a handful of works have employed EEG as an objective biomarker for CL estimation, and (b) very few among them computationally estimate CL in a coarse-grained manner (\textit{i.e.,} under a classification setting). In this regard, we (a)  substantially improve on the classification performance achieved in~\cite{Bilalpur} by utilizing multiple EEG descriptors and learning architectures, and (b) also show that similar EEG encodings are induced by psychoacoustic parameters eliciting similar CL. Notably, we achieve these results on data recorded via the commercial and portable \textit{Emotiv} device, known to have a low Signal-to-Noise Ratio (SNR); implying that effective machine learning can enable inference from noisy data, facilitating large-scale studies.

\section{Experimental Design \& Protocol}\label{Sec:ED}
This section details the data collection method employed for our study, stimuli and experimental protocol. Interested readers can refer to our prior study~\cite{Bilalpur} for more details.

\subsection{Participants}
Our study involved 20 participants whose average age was 28.9±4.9 years. None had formal music training, and participated following informed consent.

\begin{figure*}[t]
\centering
{\includegraphics[scale=0.60]{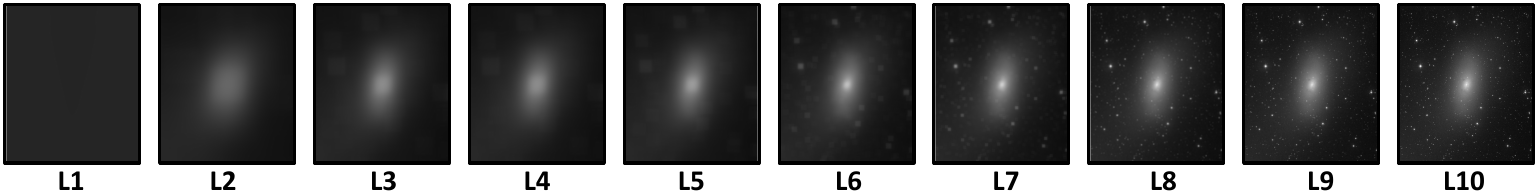}}%
\caption{ Images of the \textit{Messier 110} galaxy corresponding to focus levels ranging from 1--10 are presented from left to right.}
\label{Psycho}
\end{figure*}

\subsection{Stimuli (Psychoacoustic parameters)}
The following psychoacoustic parameters derived from~\cite{9} were examined to convey 10 focus-levels of an astronomical image shown in Figure~\ref{Psycho}. 

\subsubsection{Noise}
To represent 10-levels of varying focus with noise, a combination of a pure tone at 1000 Hz with broadband white noise was generated. A fully focused image is indicated by 100\% pure tone, while a completely defocused image is indicated by 100\% noise.

\subsubsection{Pitch}
To represent image focus levels with pitch, sinusoidal tones on the C-major scale starting from C4 (261.63 Hz) to E6 (1318.51 Hz) were generated. The underlying assumption is that higher frequencies can indicate higher levels of focus, and lower tones indicating lower focus levels.

\subsubsection{Roughness (Rough)}
To denote image focus levels with roughness, an amplitude modulation ranging from 0 to 70 Hz (specifically, 0, 2, 4, 7, 11, 16, 23, 34, 49, and 70 Hz) was applied to a pure tone at 1000 Hz. A pure tone without any amplitude modulation is considered an indicator of a focused image, while a pure tone modulated at 70 Hz indicates a completely defocused image.

\subsubsection{Noise and Rough Combined (AudioComb)}
Image focus levels were represented with a combination of roughness and noise, where the highest level of focus was indicated by a pure tone at 1000 Hz, and the lowest focus level indicated by 100\% broadband white noise modulated at 70 Hz.

\subsubsection{Image Only (Visual)}
Finally, the original astronomical images were used to benchmark their perception against acoustic parameters. Varying focus levels were generated by applying Gaussian filters of varying kernel sizes, where a clear image represents the highest level of focus (level 10).

\subsubsection{Visual and AudioComb Combined (VisualComb)} 
Astronomical images are combined with \emph{AudioComb} (\emph{i.e.}, noise and roughness combined) to investigate whether the perception of image focus is enhanced by the fusion of auditory and visual modalities.

\subsection{Experimental Protocol}
\label{E_Proto}
The experiment comprised two sessions, with each session further divided into six sub-sessions. Four of these sub-sessions involved acoustic parameters (AudioComb, Noise, Pitch, and Roughness), while two involved visual, and visual plus acoustic parameters (VisualComb). In each sub-sessions, we performed 30 stimulus presentations or \textit{trials}, where the ten stimuli were each played thrice for two seconds in a random order. The random presentation of stimuli eliminates the possibility of any pattern-learning artifacts, and ensures that each trial requires an independent user evaluation. The two main sessions involved \emph{immediate} and \emph{compared} recall as described below.

\begin{itemize}
    \item \textbf{Immediate Recall (IR):} IR testing is commonly applied for evaluating short-term memory performance over a few seconds to minutes. In the IR session, participants were instructed to immediately rate the focus level of the presented stimulus on a scale of 1--10. 
    \item \textbf{Compared Recall (CR):} The CR session mimics the Rapid Serial Visual Presentation (RSVP) protocol~\cite{rsvp}, which is widely applied in psychophysical studies for measuring perception and attention. CR is designed to elicit perceptual comparisons among stimuli presented in contiguous trials. In the CR session, participants were presented with stimuli in random order, and instructed to detect a pre-specified focus level by selecting select a \emph{Yes} radio button.
\end{itemize}

\begin{figure*}[h]
\centering
{\includegraphics[scale=0.54]{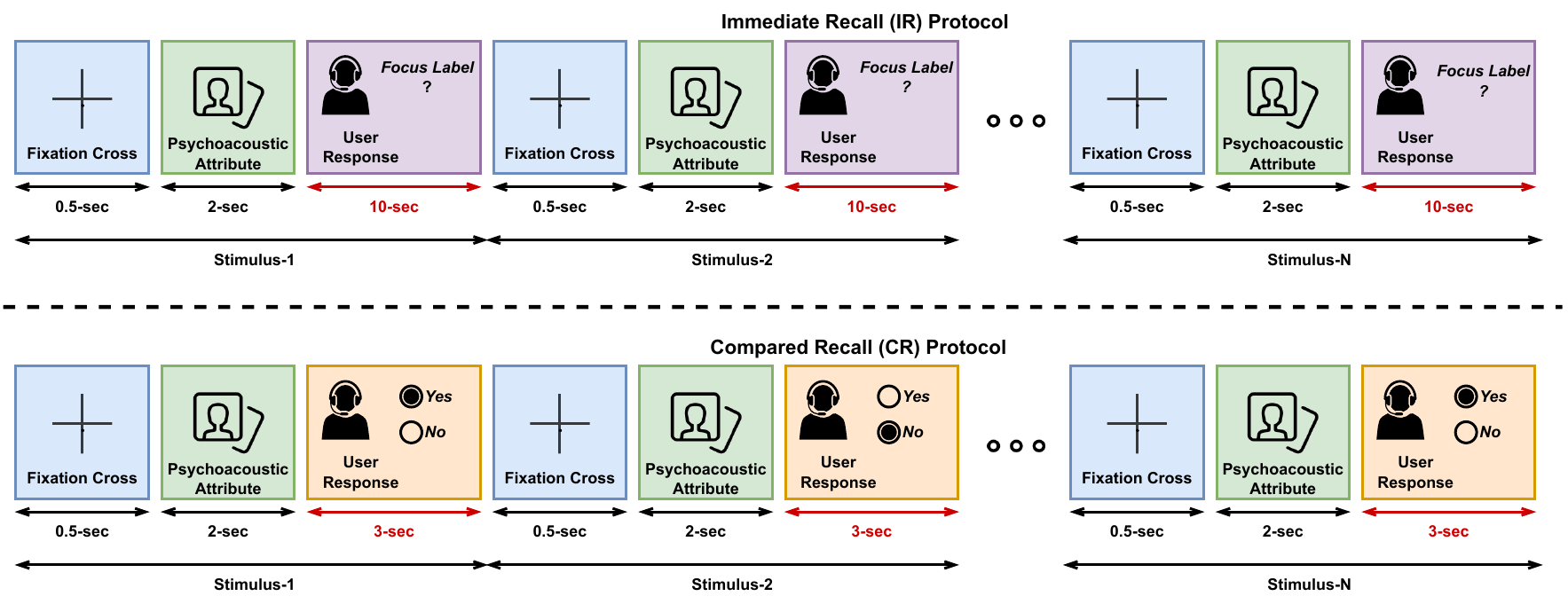}}
\caption{Overview of the Immediate Recall (IR) and Compared Recall (CR) protocols. Refer to section~\ref{E_Proto} for details.}
\label{ICR}
\end{figure*}

Each IR stimulus was presented for 2s and was preceded by a 500ms fixation cross following which users rated their perceived focus level on a 1-10 scale. A 10s countdown appeared on the screen during the judgment task, and if the user did not respond within this time, the next stimulus was automatically presented. The main difference between the IR vs CR sessions was that a 3s response time was provided for CR, as against a 10s response time for IR.

On completing each sub-session, participants rated their expended cognitive load using a subset of the NASA-TLX parameters~\cite{tlx}. These parameters included \textit{effort} (effort required to complete the task), \textit{mental demand} (level of stress while performing the task), and \textit{frustration} (the degree of frustration experienced during the task). Participants rates these attributes on a scale of 1--4, with 4 indicating the highest CL level. It is assumed that each of the above attributes complementarily add to the CL.

A coded stimulus representation (Type 1-6) was applied, avoiding scientific terms to ensure that users understood their tasks well. Additionally, a practice session was given to participants before the experiment for familiarization. This ensures they receive sufficient training and understand the stimuli and task requirements. Furthermore, the two sessions were recorded separately with a break in-between to reduce the impact of fatigue on user performance. The IR and CR protocols are illustrated in Figure~\ref{ICR}.

\section{EEG Dataset \& Feature Extraction}\label{Sec:EEG_FE}

\subsection{EEG Dataset}
User EEG responses in the IR and CR sessions were recorded via a 14-channel consumer-grade \textit{Emotiv Epoc} device. The sampling rate was set to 128Hz. While collecting EEG data with consumer-grade equipment, noise and artifacts are commonly encountered. To clean the recorded data, we applied the following preprocessing steps. Recordings were segmented into epochs, with each epoch denoting an episode involving 0.5s fixation-cross plus 2s stimulus presentation. Elimination of the DC offset in neural activity was achieved by analyzing EEG responses for the fixation-cross. A sixth-order Butterworth bandpass filter~\cite{butter}, configured to work in 0.1-45 Hz range, was then applied to remove artifacts related to muscle movements. Independent component analysis~\cite{ica} based filtering was applied to remove potential artifacts. Lastly, a detailed visual inspection was performed to ensure that the data are free of highly noisy epochs, and are of the highest quality. 

\subsection{Feature Extraction}\label{feats}
The following features were extracted from the preprocessed EEG recordings.

\begin{table}[b]
\centering
\caption{Distribution of \emph{high}, \emph{low} CL EEG epochs in IR \& CR settings.}
\label{odist}
\begin{tabular}{c|ccc}
Setting & Total Epochs & High CL & Low CL \\
\hline
IR & 2708 & 1819 & 889 \\
CR & 258 & 192 & 66
\end{tabular}
\end{table}

\subsubsection{Power Spectral Density Vectors}
\label{PSDV}
Power spectral density (PSD)~\cite{psd} provides insights into power distribution across various EEG bands and can be applied to analyze brain activities. Primary spectral bands of interest are $\delta$ (1-4 Hz), $\theta$ (4-8 Hz), $\alpha$ (8-12 Hz), $\beta$ (12-30 Hz), and $\gamma$ ($\geq$30 Hz), which have been extensively investigated in the literature~\cite{eeg1,eeg2,eeg3}. For our baseline analysis, we computed the PSD estimates of the above frequency bands by applying a sixth-order Butterworth bandpass filter~\cite{butter} and using the Welch method, and concatenated them to obtain a vector representation.

\begin{figure}[h]
\centering
{\includegraphics[scale=0.55]{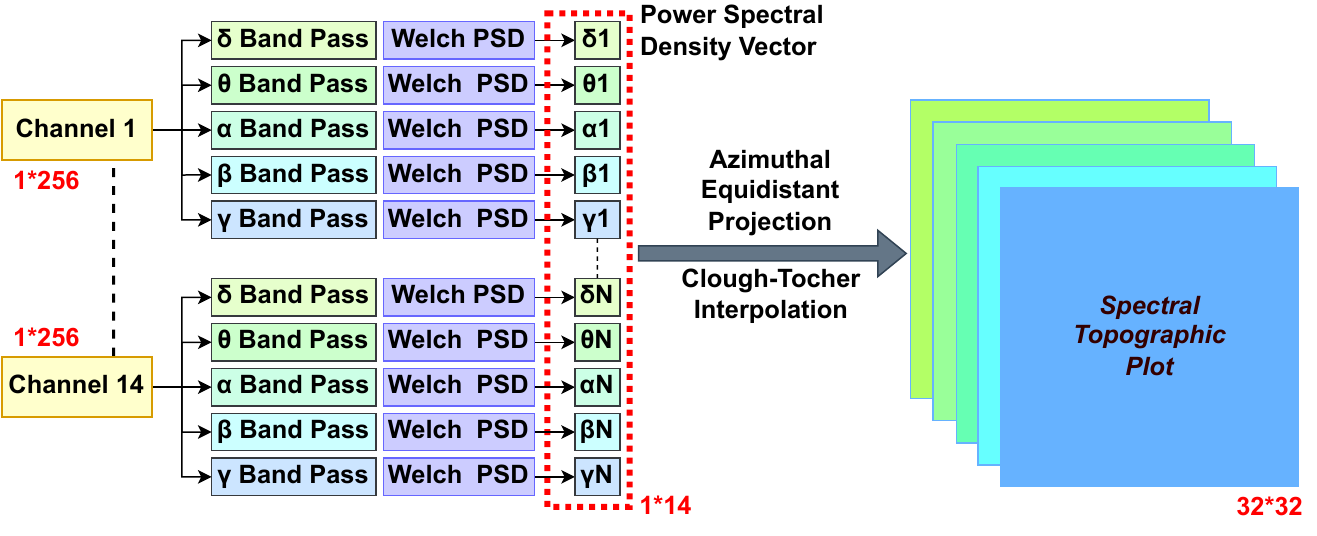}}
\caption{Power spectral density vectors and spectral topographic plots. Refer to sections~\ref{PSDV} and~\ref{STP} for details.}
\label{psd_topo_cnns1}
\end{figure}

\subsubsection{Spectral Topographic Plots}
\label{STP}
As PSD features do not capture the spatial structure of EEG data, Bashivan \textit{et al.} \cite{Pouya_iclr} proposed spectral topographic plots, which preserve both spectral and spatial information by transforming spectral power estimates into topographic images. These images display the spatial distribution of neural activity across the cortex, with each electrode representing the activity map of a specific frequency band. To generate these maps, we first projected PSDs for the 14 electrodes of the \textit{Emotiv} EEG sensor onto a 2D surface using an azimuthal equidistant projection~\cite{azi}. This projection preserves the distance between the center and all other points. We then applied Clough-Tocher \cite{Clough-Tocher} interpolation to interpolate PSD estimates between electrodes. Figure~\ref{psd_topo_cnns1} outlines this approach.

\subsubsection{Channel-wise STFT Spectrograms}
In addition to spectral topographic plots, we generated short-term Fourier transformation (STFT)~\cite{stft} based spectrogram plots. Applying STFT to EEG signals, we obtain images that represent temporal spectral changes. This representation reduces the impact of signal noise and enables us to extract better features for analyzing brain activity. 

\section{User Response Analysis}\label{Sec:ERA}
We begin our investigation on CL modeling via explicit user data, namely, data-to-parameter mapping accuracy and NASA-TLX ratings. 

\subsection{Mapping Accuracy}\label{mapp_acc}
The data-to-parameter mapping accuracy signifies how well users can identify the focus-level via each psychoacoustic parameter. Mapping accuracies for the IR and CR settings are detailed in Table~\ref{ERA}.
\begin{table}[!bp]
\centering
\caption{Data mapping accuracies for the different parameters.}
\label{ERA}
\begin{tabular}{l|cc}
Parameter & IR & CR\\
\hline
Noise & 0.64$\pm$0.10 &  0.67$\pm$0.26 \\
Pitch & 0.57$\pm$0.11 & 0.54$\pm$0.30 \\ 
Rough & 0.48$\pm$0.19 & 0.63$\pm$0.35 \\
AudioComb & 0.71$\pm$0.18 & 0.64$\pm$0.32\\
Visual & 0.78$\pm$0.18 & \textbf{0.84$\pm$0.22}\\
VisualComb & \textbf{0.83$\pm$0.11} & 0.82$\pm$0.33\\ 
\end{tabular}
\end{table}

\begin{itemize}
    \item For the IR setting, the highest mapping accuracy was achieved by~\emph{VisualComb}. This conveys that combining visual cues alongside acoustic parameters enhances user's stimulus perception. We also observe that the \emph{Visual} parameter achieves better mapping accuracy than all acoustic parameters. On closer examination of individual acoustic parameters, \emph{AudioComb} attains the highest mapping accuracy consistent with prior findings in~\cite{9}. However, an exception is that \emph{Pitch} exhibits better mapping accuracy than \emph{Rough}.

\item We observed that the highest mapping accuracies were achieved with \emph{AudioComb} and \emph{Noise}, whereas \emph{Pitch} exhibited the lowest accuracy. Mapping accuracies for \emph{Rough} and \emph{AudioComb} were quite comparable. Furthermore, we noted a higher accuracy variance in the CR setting compared to IR, revealing that the CR setting was more challenging.
\end{itemize}

\subsection{NASA-TLX rating-based CL classifiation}
We considered the mean score over the three NASA-TLX parameters (effort, mental demand and frustration) to benchmark overall CL. Given the 0--4 rating scale, we applied a threshold of 2 to categorize a parameter as inducing \emph{(high/low)} CL. In this way, each EEG epoch was labeled as either \emph{low} or \emph{high} cognitive load. Table~\ref{odist} lists the distribution of high/low CL EEG epochs for the IR and CR settings.



\section{EEG-based CL prediction} 

Upon labeling each EEG epoch based on NASA-TLX ratings, we proceeded to predict induced CL from EEG signals employing the following classifiers.

\subsection{Traditional Machine Learning Methods}
\label{TMLA}
As baselines, we assigned PSD vectors (Sec.~\ref{feats}) to high/low CL labels employing traditional machine learning classifiers listed below.

\begin{itemize}
    \item \emph{\textbf{Gaussian Naive Bayes (GNB)}} is a generative model designed to learn priors and conditionals to estimate posteriod likelihoods. It assumes that the training data are derived from a multidimensional Gaussian, and class-conditional feature independence.
    
    \item \emph{\textbf{Linear Discriminant Analysis (LDA)}} is a discriminative model that identifies the best linear feature combination for separating multiple classes. Its aim is to reduce data dimensionality while retaining class-discriminatory information. For a $C-$class task, LDA derives a $C-1$ dimensional sub-space maximizing class separation.
    
    \item \emph{\textbf{Support Vector Machine (SVM)}} is a maximal margin classifier, seeking a linear boundary for data separation in the original, or a transformed high-dimensional feature space.
\end{itemize}

\begin{figure*}[h]
\centering
{\includegraphics[angle=90, scale = 0.90]{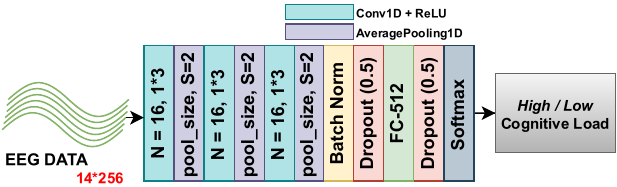}}
\caption{Temporal-CNN Architecture. Refer to section~\ref{T_CNN} for details.}
\label{CNN1D}
\end{figure*}

\begin{figure*}[t]
\centering
\includegraphics[scale=0.85]{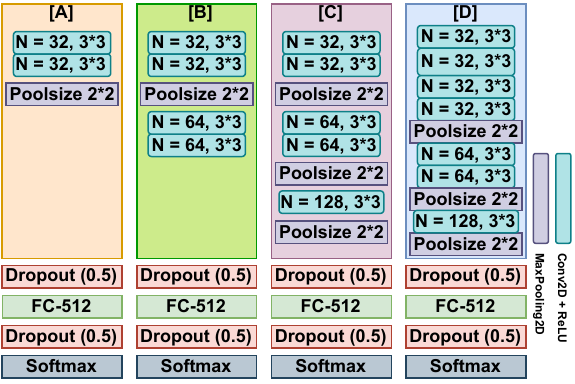}
\caption{Spatial CNN Architectures. Refer to section~\ref{S_CNN} for details.}
\label{psd_topo_cnns}
\end{figure*}

\subsection{Temporal CNN}
\label{T_CNN}
On time-series EEG data, we applied a 1D-CNN to retrieve features encoding the sequential information~\cite{1_dcnn}. The 1D-CNN comprised three stacked blocks (Figure~\ref{CNN1D}), where each block consists of alternating 1-D convolution and max-pooling layers. The output of the convolutional blocks is flattened and routed through a batch normalization layer~\cite{batchN}. To prevent overfitting, we apply a 50\% dropout layer. 

In addition to the 1D CNN, we also evaluated EEGNet~\cite{eegnet} designed for efficient EEG signal classification. EEGNet comprises depth-wise separable convolution layers which divide the convolution into spatial and channel-wise stages, which enables learning spatio-temporal features with fewer parameters~\cite{eegnet}.

\subsection{Spatial CNN}
\label{S_CNN}
We adopted the CNN architectures detailed in~\cite{Pouya_iclr} to estimate CL using spectral images. These architectures consist of convolution layers followed by a max pool operation to downsample feature maps. Rectified linear unit (ReLU)~\cite{relu} activation functions are applied to introduce non-linearity, while a dropout layer is added after the fully connected layers to avoid overfitting. Figure~\ref{psd_topo_cnns} details these CNN architectures.

\begin{figure*}[t]
\centering
\includegraphics[scale = 0.85]{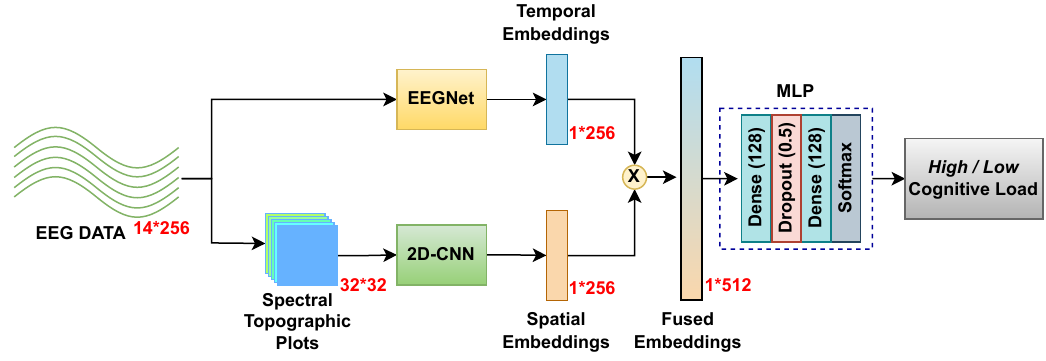}
\caption{Late Fusion of Temporal \& Spatial encodings. Refer to section~\ref{late_fusion} for details.}
\label{fusion}
\end{figure*}

\subsection{Fusion of Temporal \& Spatial CNN}
\label{late_fusion}
As the EEGNet and the topoplot-based 2D-CNN architectures are respectively designed to effectively learn temporal and spatial EEG patterns, we explored if fusing these representations enables efficient modeling of spatio-temporal EEG correlates of CL. To this end, we fused the features learned from the best-performing temporal and spatial CNNs, allowing us to capture temporal dependencies, spatial relationships, and their interactions. We extracted features derived from the fully-connected (fc) temporal and spatial CNN layers (Figure~\ref{fusion}), each of which comprise 256 neurons, and concatenating their outputs results in a 512-dimensional descriptor. We then employed a multi-layered perceptron (MLP) network comprising two dense layers with 128 neurons each with Relu activations. A 50\% dropout was applied between the two dense layers, followed by softmax classification.  

\subsection{Model Training \& Validation Procedure}
To perform a rigorous evaluation of the above models performance and ensure their generalizability, we validated all models using ten iterations of 5-fold cross-validation. We choose the F1 score as our performance metric due to our class-imbalanced data.

\begin{table*}[b]
\centering
\caption{Cognitive load classification for the IR and CR settings.}
\label{clc_r}
\begin{tabular}{@{}l|ccc|lccc@{}}
\multicolumn{1}{c}{} & \multicolumn{3}{c}{IR} & \multicolumn{3}{c}{CR} \\ \midrule
Method & Precision & Recall & F1 Score & Precision & Recall & F1 Score  \\ 

\midrule
GNB & 0.52$\pm$0.08  & 0.42$\pm$0.08  & 0.41$\pm$0.09  & 0.75$\pm$0.08  & 0.50$\pm$0.09  & 0.50$\pm$0.09 \\
LDA & 0.54$\pm$0.07  & 0.58$\pm$0.07  & 0.53$\pm$0.08  & 0.62$\pm$0.06  & 0.59$\pm$0.06  & 0.60$\pm$0.07 \\
L-SVM  & 0.56$\pm$0.05  & 0.58$\pm$0.04  & 0.55$\pm$0.04  & 0.64$\pm$0.06  & 0.69$\pm$0.06  & 0.66$\pm$0.06 \\
R-SVM  & 0.56$\pm$0.01  & 0.63$\pm$0.01  & 0.55$\pm$0.01  & 0.55$\pm$0.02  & 0.74$\pm$0.02  & 0.64$\pm$0.02 \\
\midrule
1D-CNN & 0.64$\pm$0.05  & 0.66$\pm$0.06  & 0.64$\pm$0.04  & 0.65$\pm$0.07  & 0.68$\pm$0.06  & 0.66$\pm$0.07 \\
EEGNet & 0.66$\pm$0.04  & 0.67$\pm$0.04  & 0.66$\pm$0.05  & 0.67$\pm$0.08  & 0.70$\pm$0.06  & 0.67$\pm$0.06 \\
\midrule
Topographic Plots (A) & 0.72$\pm$0.03  & 0.73$\pm$0.03 & 0.72$\pm$0.03  & 0.66$\pm$0.05  & 0.68$\pm$0.05  & 0.66$\pm$0.05 \\
Topographic Plots (B) & 0.69$\pm$0.03  & 0.70$\pm$0.03  & 0.69$\pm$0.03  & 0.64$\pm$0.05  & 0.68$\pm$0.05  & 0.65$\pm$0.05 \\
Topographic Plots (C) & 0.69$\pm$0.03  & 0.70$\pm$0.03  & 0.69$\pm$0.03  & 0.67$\pm$0.05  & 0.62$\pm$0.04  & 0.64$\pm$0.05 \\
Topographic Plots (D) & 0.60$\pm$0.10  & 0.68$\pm$0.03  & 0.61$\pm$0.05  & 0.60$\pm$0.06 & 0.69$\pm$0.05  & 0.63$\pm$0.03 \\
STFT Spectrogram (A) & 0.70$\pm$0.02  & 0.71$\pm$0.02  & 0.70$\pm$0.02  & 0.61$\pm$0.08  & 0.72$\pm$0.03  & 0.64$\pm$0.03 \\
STFT Spectrogram (B) & 0.71$\pm$0.02  & 0.72$\pm$0.02  & 0.70$\pm$0.02  & 0.58$\pm$0.06  & 0.73$\pm$0.02  & 0.63$\pm$0.02 \\
STFT Spectrogram (C) & 0.70$\pm$0.02  & 0.71$\pm$0.02  & 0.70$\pm$0.02  & 0.59$\pm$0.06  & 0.73$\pm$0.02  & 0.64$\pm$0.02 \\
STFT Spectrogram (D) & 0.70$\pm$0.02  & 0.71$\pm$0.02  & 0.70$\pm$0.02  & 0.60$\pm$0.07  & 0.69$\pm$0.08  & 0.63$\pm$0.07 \\
\midrule
Spatio-Temporal+MLP & \textbf{0.98$\pm$0.01}  & \textbf{0.98$\pm$0.01}  & \textbf{0.98$\pm$0.01}  & \textbf{0.96$\pm$0.02}  & \textbf{0.96$\pm$0.02}  & \textbf{0.96$\pm$0.02} \\

\end{tabular}
\end{table*}

\begin{figure*}[h] 
    \centering
    \subfigure{\includegraphics[width=0.33\textwidth]{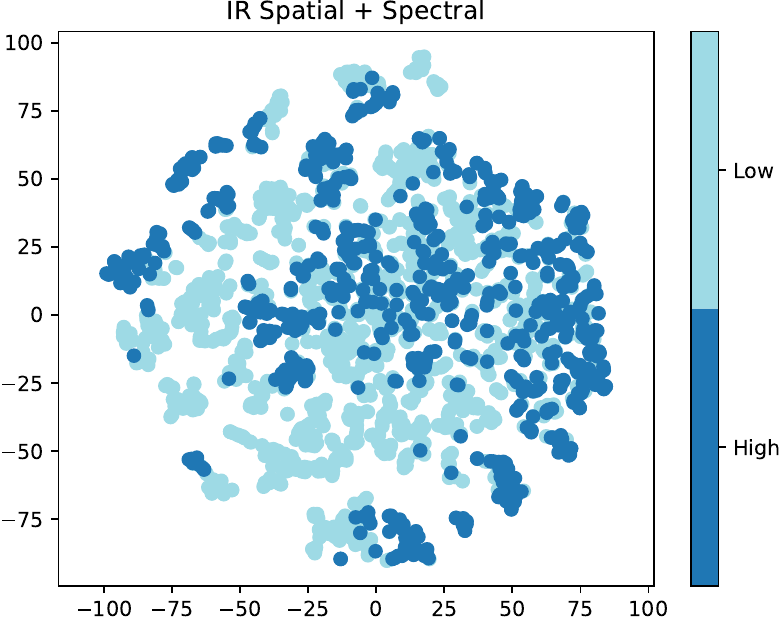}} 
    \subfigure{\includegraphics[width=0.33\textwidth]{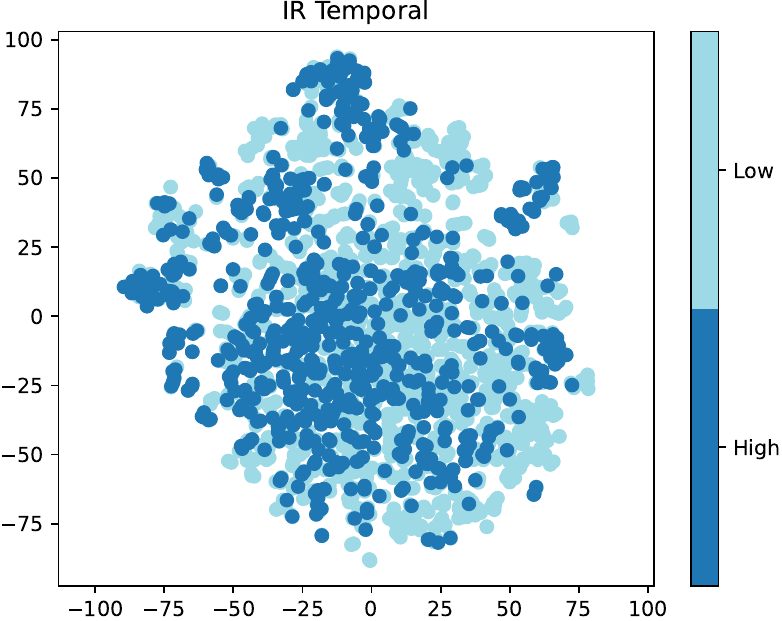}} 
    \subfigure{\includegraphics[width=0.33\textwidth]{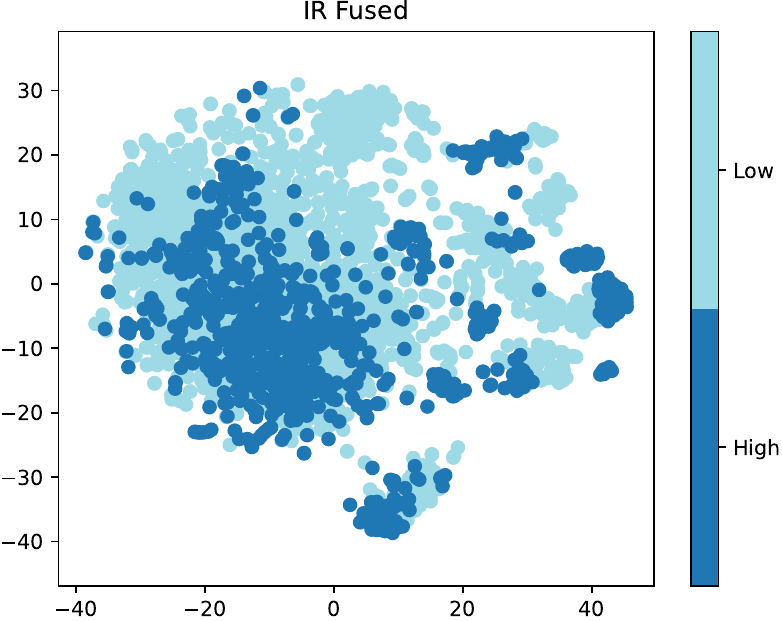}}
    \subfigure{\includegraphics[width=0.33\textwidth]{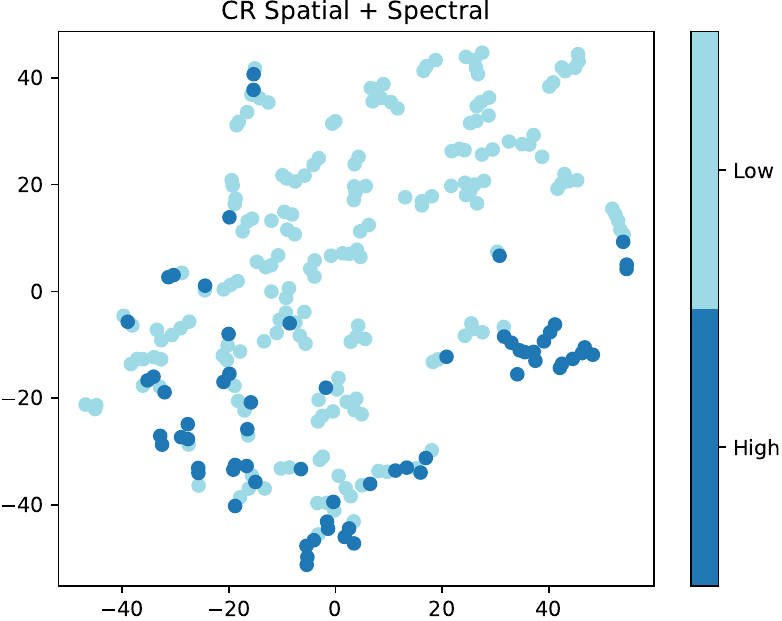}} 
    \subfigure{\includegraphics[width=0.33\textwidth]{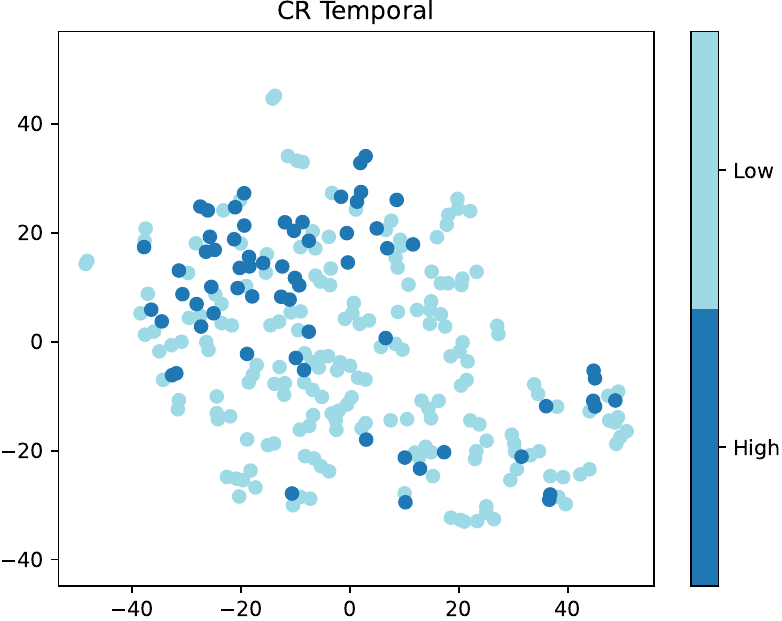}} 
    \subfigure{\includegraphics[width=0.33\textwidth]{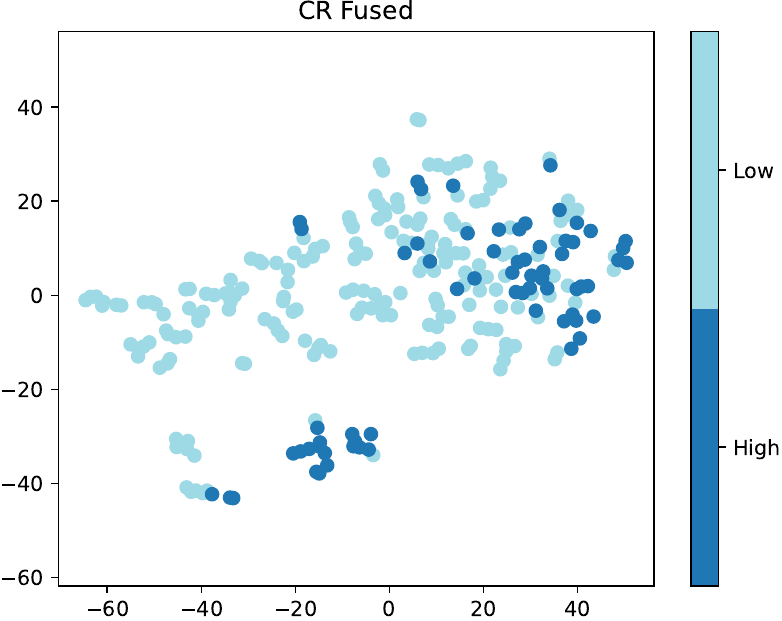}}   
    \caption{\textit{t}-SNE plots of the spatial, temporal and fused embeddings in the IR (top) and CR (bottom) settings.}
    \label{umap} \vspace{-2mm}
\end{figure*}

\subsection{Experimental Results}
CL classification scores achieved with the above models are depicted in Table \ref{clc_r}, and discussed as follows.

\subsubsection{Traditional ML Classification}
ML classifiers achieved near-to or better-than-chance CL classification, with GNB performing worst and RBF-SVM performing best. Given that optimal CL estimation in~\cite{Bilalpur} is achieved via a CNN, we now examine performance with deep learning architectures.

\subsubsection{Temporal CNN}
We observe 1D-CNN work better than traditional ML classification algorithms as it provides better feature representation of EEG signals. For both IR and CR, we observe that EEGNet performs slightly better than the 3-layered 1D-CNN in terms of precision, recall, and F1 score. This improvement can be attributed to the presence of depthwise separable convolutional layers in EEGNet, enabling it to learn spatio-temporal EEG patterns using fewer parameters.

\subsubsection{Spatial CNN}
We explored the efficacy of two EEG representations, namely spectral topographic plots and STFT spectrograms (Sec.~\ref{feats}), with four 2D-CNN architectures depicted in Fig.~\ref{psd_topo_cnns}. Table~\ref{clc_r} conveys that spectral topographic plots tend to slightly outperform STFT spectrograms in the CR setting, while mixed results are noted in the IR setting. In the IR setting, architecture \textit{A} achieves optimal performance with topographic plots, consistent with prior findings~\cite{gs_icmi_22}; topographic maps denote simpler features, requiring less complex processing  than spectrograms. In the CR setting, architecture \emph{A} again achieves the best precision and F1-score, while architecture \emph{D} achieves the best recall. Overall, a higher variance in classification measures is noted in the CR setting as compared to IR. With STFT spectrograms, we observe minimal variance in performance across architectures, indicating its robustness to changes in network architecture.

\subsubsection{Spatio-Temporal Fusion CNN}
It is evident from Table~\ref{clc_r} that the fused representations optimally capture CL-related EEG correlates as compared to the spatial or temporal counterparts. Much superior and close-to-ceiling classification metrics are achieved with the fusion+MLP framework (Fig.~\ref{fusion}) for both the IR and CR settings, conveying the effectiveness of the fused embeddings. This argument is further validated via Figure~\ref{sia_arch}, which plots the \textit{t}-distributed stochastic neighbor embeddings (\textit{t}-SNE) for the temporal, spectral, and fused descriptors in both the IR and CR settings. There is much greater inter-class overlap with spatial and temporal embeddings, as compared to the fused embeddings, which explains optimal CL estimation achieved with spatio-temporal fusion.

\section{EEG-based Categorization of Focus-levels}
\label{Sec:FLD}
Following on from Sec.~\ref{mapp_acc}, we examined detection sensitivity differences across focus-levels induced under multiple psychoacoustic parameters in the IR setting. To begin with, one can hypothesize that the extremum focus-levels (\eg, 1, 2, 9, 10) are relatively easier to detect than the intermediate levels. This is adequately reflected in Table~\ref{map_acc_fl}, which presents the data-to-sound mapping accuracy per focus-level. Consequently, the process of perceiving extremum vs intermediate focus-levels should also induce distinct neural encodings. To validate this hypothesis, we fed PSD vectors corresponding to the EEG responses for \textit{easy} vs \textit{difficult}-to-detect focus-levels to machine learning classifiers.

\subsection{Experimental Results}
\subsubsection{Mapping Accuracy}
\label{MA_D}
Table~\ref{map_acc_fl} summarizes acoustic parameter-induced focus-level detection accuracies in the IR setting. Similar sensitivity trends were noted across parameters with the exception of \emph{Roughness}, which induced low detection accuracies across levels. Therefore, \emph{Rough} data were ignored for analyses, and EEG-based \textit{easy} vs \textit{difficult}-to-detect focus-level categorization results are presented in Table~\ref{map_clf}. 

\subsubsection{Classification Results}\label{NC_D}
Empirical results in Table~\ref{map_clf} show that better-than-chance classification of the extremum (easy-to-detect) and intermediate (difficult-to-detect) focus-levels is achievable with classical machine learning methods. Similarities between Tables~\ref{ERA} and~\ref{map_clf} can also be noted as the two best parameters in Table~\ref{ERA} (Visual and VisualComb) consistently generate the highest F1-scores across classifiers in Table~\ref{map_clf}. Overall, the GNB and L-SVM achieve lower F1-scores than LDA and R-SVM. R-SVM achieves optimal F1-scores across parameters, conveying a non-linear boundary optimally separates \textit{extremum} vs \textit{intermediate} samples in the PSD feature space. 

\section{Perceptual Similarity Among Parameters}\label{Sec:Per_Sim}
\label{SSPA_}
\begin{table*}[t]
\caption{Mapping Accuracy per focus level.}\vspace{-2mm}
\label{map_acc_fl}
\centering
\begin{tabular}{l|cccccccccc}
Parameter & L1 & L2 & L3 & L4 & L5 & L6 & L7 & L8 & L9 & L10\\
\hline
Noise & \textbf{0.82} & 0.52 & 0.45 & 0.19 & 0.16 & 0.17 & 0.28 & 0.22 & 0.33 & 0.63\\
Pitch & 0.39 & \textbf{0.53} & 0.24 & 0.20 & 0.23 & 0.23 & 0.21 & 0.25 & 0.2  & 0.23\\ 
Rough & 0.27 & 0.16 & 0.16 & 0.15 & 0.09 & 0.15 & 0.17 & 0.2  & 0.28 & \textbf{0.44} \\
AudioComb & \textbf{0.88} & 0.46 & 0.40  & 0.18 & 0.15 & 0.19 & 0.22 & 0.30 & 0.38 & 0.65\\
Visual & 0.85 & \textbf{0.87} & 0.59 & 0.34 & 0.14 & 0.09 & 0.21 & 0.4 & 0.45 & 0.78\\
VisualComb & \textbf{0.97} & 0.78 & 0.41 & 0.24 & 0.23 & 0.25 & 0.13 & 0.46 & 0.55 & 0.85\\
\end{tabular}
\end{table*}

\begin{table*}[t]
\centering
\caption{F1-scores for EEG-based classification of extremum vs intermediate focus-levels for different parameters.}
\label{map_clf}\vspace{-2mm}
\centering
\begin{tabular}{l|cccccc}
 & Noise & Pitch & AudioComb & Visual & VisualComb\\
 \hline
GNB & 0.33$\pm$0.02 & 0.35$\pm$0.03 & 0.36$\pm$0.07 & 0.64$\pm$0.04 & 0.41$\pm$0.03 \\
LDA & 0.67$\pm$0.02 & 0.65$\pm$0.02  & 0.68$\pm$0.02 & 0.68$\pm$0.01 & 0.67$\pm$0.01 \\
L-SVM & 0.54$\pm$0.09 & 0.55$\pm$0.10  & 0.53$\pm$0.10 & 0.54$\pm$0.12 & 0.56$\pm$0.10 \\
R-SVM & \textbf{0.69$\pm$0.00} & \textbf{0.69$\pm$0.00} & \textbf{0.69$\pm$0.00} & \textbf{0.69$\pm$0.00} & \textbf{0.70$\pm$0.00} \\
\end{tabular}
\end{table*}

While prior sections reveal that distinctive EEG signatures enable categorization of (a) \textit{low} vs \textit{high} CL-inducing trials, and (b) \textit{extremum} vs \textit{intermediate} focus-levels, this section explores EEG-based \textit{perceptual similarity} estimation for parameter pairs, \ie, the possibility of estimating whether a pair of parameters are \textit{similar} or \textit{dissimilar} in terms of the cognitive load they induce. Consistent with~\cite{Bilalpur}, we identify \emph{Noise}, \emph{Pitch}, \emph{Rough}, and \emph{AudioComb} as \textit{high} CL parameters, while \emph{Visual} and \emph{VisualComb} are identified as low CL-inducing. For perceptual similarity estimation, we utilize a convolutional Siamese network and hypothesize proximal latent EEG embeddings for similar parameters. 

\begin{figure}[!h]
\centering
\includegraphics[scale = 0.60]{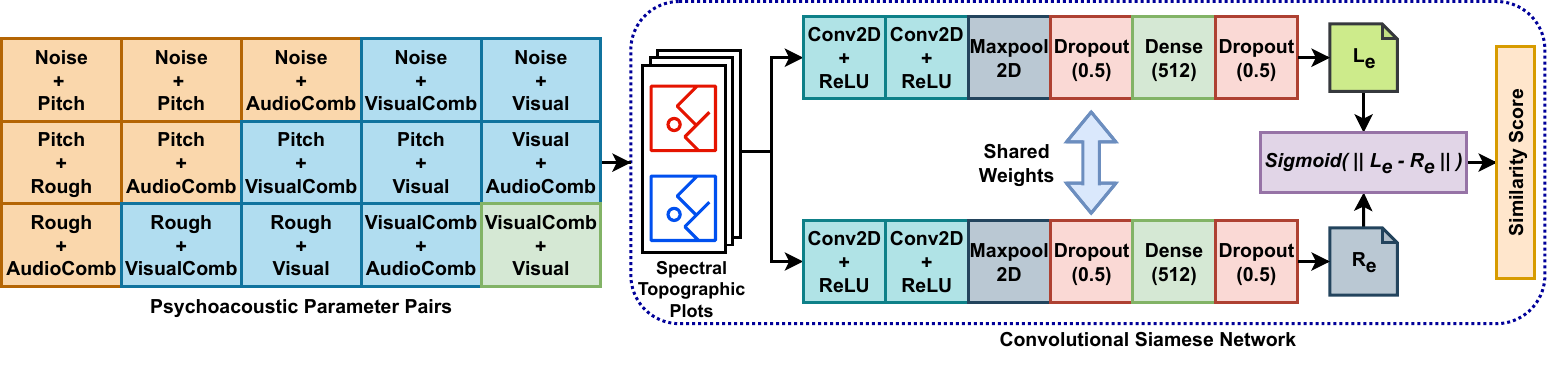}\vspace{-2mm}
\caption{Perceptual similarity estimation among psychoacoustic parameters via convolutional Siamese network.}
\label{sia_arch}
\end{figure}

\subsection{Convolutional Siamese Network}
The Siamese network employed for perceptual similarity estimation comprises twin parallel CNNs with shared weights~\cite{sia1,sia2}. These parallel CNNs act as encoder blocks, mapping the input onto a latent feature space. When these CNNs are identical, similar samples are mapped to the same latent subspace and the relative distances between these embeddings can then provide the notion of a \textit{similarity score}. The choice of the \textit{loss function} influences this similarity score. We apply \textit{contrastive loss}~\cite{contra}, where positive pairs are formed by combining samples with identical CL labels, and negative pairs with sample pairs having different CL labels. We compute the $\ell_2$ norm between latent pair embeddings, followed by the sigmoid~\cite{sigmoid} function to estimate the similarity score. Our objective is to train the Siamese network to minimize the distance between encodings for positive pairs, and maximize distance for negative pairs. The Siamese network architecture for similarity estimation is illustrated in Figure~\ref{sia_arch}.

\subsection{Perceptual Similarity- Empirical Settings}
We train the Siamese network and validate its performance via ten repetitions of five-fold cross-validation. For data samples corresponding to the $^6C_2$ pairs of the six psychoacoustic parameters, we generate spectral topographic plots, and utilze the best-performing CNN architecture (A) from the cognitive load identification task to extract latent embeddings. Test results denote the $\mu \pm \sigma$ values over the 50 runs.

\subsection{Experimental Results}\label{SSPA}
From Table~\ref{sia_tab}, we observe that all the similarity estimates for all parameter pairs are in line with our expectation, with the exception of the \emph{Noise$-$AudioComb} and \emph{Rough$-$AudioComb}. Overall, these results confirm our hypothesis that similar CL-inducing parameters generate proximal latent EEG embeddings. A possible explanation for the anomalous results is that \textit{AudioComb} inherently combines \textit{Noise} and \textit{Roughness}, and the corresponding EEG embeddings may not align well in the same subspace.

\begin{table}[]
\caption{Perceptual similarity scores for psychoacoustic parameter pairs. A score close to 0.5 indicates \textit{similarity}, while 0 indicates \textit{dissimilarity} (Section \ref{SSPA}). Incorrect predictions are denoted in bold.}
\label{sia_tab}

\centering
\begin{tabular}{l|ll}
Attribute Pairs & CL Labels & Similarity Score\\
\hline
Noise-VisualComb & High-Low &  0.00$\pm$0.05 (Dissimilar)\\
Noise-Visual & High-Low &  0.00$\pm$0.00 (Dissimilar)\\
Pitch-VisualComb & High-Low &  0.00$\pm$0.00 (Dissimilar)\\
Pitch-Visual & High-Low &  0.00$\pm$0.00 (Dissimilar)\\
Rough-VisualComb & High-Low &  0.00$\pm$0.02 (Dissimilar)\\
Rough-Visual & High-Low &  0.00$\pm$0.04 (Dissimilar)\\
AudioComb-VisualComb & High-Low &  0.00$\pm$0.00 (Dissimilar)\\
AudioComb-Visual & High-Low &  0.00$\pm$0.00 (Dissimilar)\\ \hline
Noise-Pitch & High-High &  0.51$\pm$0.00 (Similar)\\
Noise-Rough & High-High &  0.49$\pm$0.00 (Similar)\\
Pitch-Rough & High-High &  0.48$\pm$0.00 (Similar)\\
Pitch-AudioComb & High-High &  0.50$\pm$0.00 (Similar)\\
VisualComb-Visual & Low-Low &  0.50$\pm$0.00 (Similar)\\ 
\textbf{Noise-AudioComb} & \textbf{High-High} &  \textbf{0.00$\pm$0.00 (Dissimilar)}\\
\textbf{Rough-AudioComb} & \textbf{High-High} &  \textbf{0.00$\pm$0.00 (Dissimilar)}\\
\end{tabular} \vspace{-2mm}
\end{table}


\section{Discussion \& Conclusion}\label{Sec:Con}
Sonification involves using non-speech audio to represent and visualize data. Good sonification design involves mapping information via signals  that are intuitive and easily comprehensible. With clear and distinct auditory signals, users can accurately perceive information without burdening their cognitive resources. This paper explores EEG-based estimation of cognitive load (CL) induced by psychoacoustic parameters like \textit{pitch}, \textit{roughness}, and \textit{noise}, and their combinations to convey the focus level of an astronomical image. We additionally demonstrate the utility of EEG embeddings to (a) differentiate between auditory mappings denoting \textit{extremum} vs \textit{intermediate} focus levels, and (b) determine acoustic parameter pairs inducing \textit{similar} vs \textit{dissimilar} cognitive load levels.

For CL estimation, empirical results confirm that (1) EEG descriptors are reliable and achieve considerably better-than-chance CL categorization, and (2) combining the temporal and spatial representation via a multi-layer perceptron enables precise CL categorization. Therefore, EEG can serve an authentic cognitive modality to benchmark the utility of a psychoacoustic parameter for sonification. Our experiments further reveal that discriminative EEG embeddings are obtained as users perceive extremum vs intermediate focus-levels via acoustic parameters, and that extremum levels are more precisely detected by users than intermediate ones; this observation emphasizes the need to consider alternate sonification parameters to achieve enhanced clarity. Finally, a convolutional Siamese network fed with spectral EEG descriptors can isolate parameter pairs that induce similar and dissimilar cognitive load levels. 

A limitation of this work is that it totally examines six visualization parameters for conveying the focus-level of an astronomical image. Future work will examine other parameters like \textit{loudness}, \textit{sharpness}, and \textit{fluctuation strength} to this end. We will also evaluate the potential of psychoacoustic parameters to convey properties like image color and contrast, and correlations among them. While PSD descriptors are popular for EEG-based classification, future will also explore alternative EEG features such as event-related potentials or phase coherence for their discriminability. Finally, subject-independent EEG analysis will also be studied to enable generalizable cognitive load estimation across individuals.
\bibliographystyle{ACM-Reference-Format}
\bibliography{main}

\end{document}